\documentstyle[prb,aps,twocolumn, floats]{revtex}
\newcommand{\be}{\begin{equation}}
\newcommand{\ee}{\end{equation}}
\input psfig.sty
\begin{document}
\title{Two-hole dynamics in spin ladders}
\author{Christoph Jurecka and Wolfram Brenig}

\address{Institut f\"ur Theoretische Physik,
Technische Universit\"at Braunschweig, 38106 Braunschweig, Germany\vskip.2cm}

\author{Dedicated to Professor Peter W\"olfle on the occasion
of his 60$^{\rm th}$ birthday.\vskip.3cm}

\date{\today}
\maketitle
\begin{abstract}
We present an analytic theory for the energy spectrum of a two-leg spin
ladder doped with two holes. Starting from a pseudo-fermion-bond-boson
representation of the corresponding $t_{1,2}-J_{1,2}$ Hamiltonian we apply a
diagrammatic approach adapted to the limit of strong rung coupling, which
includes both, the coupling of holes to the spin background as well
as the two-hole interactions. The two-hole spectrum is calculated and the
formation of bound states is discussed. Additionally the evolution of the
spin gap of the ladder upon doping is analyzed. A comparison with existing
exact diagonalization data is presented and good agreement is found.
\end{abstract}

\begin{section}{Introduction}
Low dimensional quantum-spin systems with local moments
arranged in ladder-type geometries have attracted considerable interest in
recent years. From a theoretical point of view spin ladders are of
interest as a bridging step from one- to two-dimensional
systems\cite{rev}. From an experimental point of view various spin ladder
materials, e.g.
SrCu$_2$O$_3$\cite{Azuma9194} or CaV$_2$O$_5$  \cite{Iwase96}, have been
analyzed, allowing to gauge theoretical findings against real systems.

Spin-$1/2$ two-leg ladders display a quantum disordered ground state
with exponentially decaying spin-correlations and a gapful magnetic
excitation spectrum. Apart from magnetic excitations
the dynamics of charge carriers doped into the quantum disordered ground
state is of
interest, in particular because of the discovery of superconductivity
\cite{Uehara96} under pressure in the spin-ladder compound
Sr$_{14-x}$Ca$_x$Cu$_{24}$O$_{41}$ \cite{Carron88}. The properties of a
single hole doped into the spin ladder have been the subject of various
numerical\cite{tsunetsugu94,tsunetsugu95,troyer96,Haas96,white97,brunner01}
and analytical studies\cite{eder98,Sushkov99,jurecka01}. Moreover, in
order to understand two-hole interactions as well as pairing
correlations several numerical investigations have been performed
focusing on various topics like the formation
of bound hole-pairs\cite{dagotto92,gazza99,siller01}, ground
state properties\cite{mueller98,rommer00} and excitation
spectrum\cite{tsunetsugu94,tsunetsugu95,troyer96,poilblanc95,dagotto98,riera99,poilblanc00},
correlation functions\cite{white97,tsutsui01} and superconducting
fluctuations\cite{hayward95,hayward96}. Regarding analytical approaches
however, only a restricted set of results is available like high order
series expansions\cite{oitmaa99}, mean-field approaches\cite{sigrist94,lee99},
effective Hamiltonian mappings\cite{laeuchli99}, perturbative
renormalization group analysis\cite{lin98}, exact solution of a related 
model\cite{bose98} and recurrent variational approach \cite{sierra98}. 
In particular, the two-hole
spectrum remains an interesting issue with many open questions.

In this work we present an analytic theory of the energy spectrum of a
two-leg spin ladder doped with two holes. Our approach allows for a
direct understanding of the relevant processes. Moreover, while showing
results in good agreement with earlier numerical
studies\cite{tsunetsugu94,tsunetsugu95,troyer96,white97,dagotto92,poilblanc95}
on small systems, our method may equally well be applied to large systems
close to the thermodynamic limit.

For the two-leg spin ladder we consider a $t-J$-model with hopping- and
exchange-integrals along the rungs and chains of fig. {\ref{fig1}},
i.e. $t_1, J_1$ and $t_2, J_2$ respectively. The Hamiltonian reads
\begin{eqnarray}
H&=&H_J+H_t \nonumber\\
H_J&=&J_1 \sum_n ({\bf S}_{1, n} {\bf S}_{2, n}-\frac{1}{4} n_{1,n}n_{2,n})  \nonumber\\ &+& 
J_2 \sum_{i,a} ({\bf S}_{a,n} {\bf S}_{a,n+1}-\frac{1}{4} n_{a,n}n_{a,n+1})
\nonumber\\ 
H_t&=&-t_1\sum_{n, \sigma} \hat{c}_{1, n, \sigma}^\dagger \hat{c}_{2, n, 
\sigma}^{\phantom\dagger}+h.c.\nonumber \\& &-t_2\sum_{n,a, \sigma}
\hat{c}_{a, n
\sigma}^\dagger \hat{c}_{a, n+1, \sigma}^{\phantom\dagger}+h.c.
\label{e1}
\end{eqnarray}
where ${\bf S}_{i,n}$ are spin-$1/2$ operators on site $i$ of rung $n$ and
$\hat{c}_{i,n, \sigma}^{(\dagger)}=[c_{i,n, \sigma} (1-n_{i,n,
-\sigma})]^{(\dagger)}$ are fermion-operators for spin
$\sigma$ and site $i,n$ projected onto the space of no double occupancy.
\begin{figure}[b]
\centerline{\psfig{file=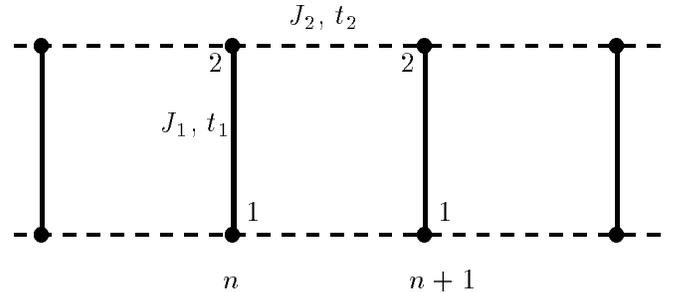,width=\linewidth}}
\caption[l]{Two-leg $t$-$J$-ladder. $n$ labels the rungs,
$1$ and $2$ denote the sites on the rung.}
\label{fig1}
\end{figure}

The paper is organized as follows. In the next section we briefly restate
the bond-operator description of the spin dynamics of the undoped ladder.
In section \ref{sec4} a mapping of the $t$-$J$ Hamiltonian in the two-hole
sector onto an interacting fermion-boson-model is described. Section
\ref{sec5} presents a diagrammatic analysis for the energy spectrum.
In section \ref{sec7} results are presented and compared to existing
numerical data. Finally conclusions will be given.
\end{section}

\begin{section}{Spin dynamics of the undoped ladder}
\label{sec3}

To describe the spin excitations of the two-leg ladder in the sector
of no holes we use the well established bond-operator representation
\cite{jurecka01,sachdev90} of dimerized spin-$1/2$ systems. Here we
briefly recapitulate this method. 
The eigenstates of the total spin on a single rung occupied
by two electrons are a singlet and three triplets. These can be created
by the bosonic bond operators $s_n^\dagger$ and $t_\alpha^\dagger$
with $\alpha=x,y,z$ acting on a vacuum $|0\rangle$ by 
\begin{eqnarray}
s_n^\dagger|0\rangle = \frac{1}{\sqrt{2}}\left(|\!\uparrow
\downarrow\rangle-|\!\downarrow\uparrow\rangle\right)_n\nonumber \\
t_{x,n}^\dagger|0\rangle = \frac{-1}{\sqrt{2}}\left(|\!\uparrow
\uparrow\rangle-|\!\downarrow\downarrow\rangle\right)_n\nonumber \\
t_{y,n}^\dagger|0\rangle = \frac{i}{\sqrt{2}}\left(|\!\uparrow
\uparrow\rangle+|\!\downarrow\downarrow\rangle\right)_n\nonumber \\
t_{z,n}^\dagger|0\rangle = \frac{1}{\sqrt{2}}\left(|\!\uparrow
\downarrow\rangle+|\!\downarrow\uparrow\rangle\right)_n
\label{e4}
\end{eqnarray}
where the first (second) entry in the kets refers to site $1$($2$) of the
rung $n$ of Fig. \ref{fig1}. On each site we have $[s,s_{\phantom{\alpha}
}^\dagger]=1$,
$[s_{\phantom{\alpha}}^{(\dagger)},t_\alpha^{(\dagger)}]=0$, and
$[t_\alpha^{\phantom{(\dagger)}},t_\beta^{\dagger}]=\delta_{\alpha
\beta}$.
The bosonic Hilbert space has to be restricted to either one singlet or
one triplet per site by the constraint
\be
s_n^\dagger s_n^{\phantom\dagger}+t_{\alpha, n}^\dagger t_{n,
\alpha}^{\phantom\dagger}=1.
\label{e5}
\ee

Expressing the spin part $H_J$ of the Hamiltonian (\ref{e1}) by the bond
operators yields an bose gas of singlets and triplets with two-particle
interactions mediated by the inter-rung coupling $J_2$.
At $J_2=0$ these interactions vanish leaving a sum of purely
local rung Hamiltonians, which lead to a product ground-state of
singlets localized on the rungs and a set of $3^N$-fold degenerate
triplets. For finite $J_2$ the inter-rung interactions can be treated
approximately by a linearized Holstein-Primakoff (LHP)
approach\cite{jurecka01,Chubokov89a,chubukov91,Starykh96a,brenig98}. The
LHP method retains spin-rotational invariance and reduces $H_J$ to a set
of three degenerate massive magnons
\be
H_J=\sum_k \omega_k \gamma_{\alpha,k}^\dagger
\gamma_{\alpha,k}^{\phantom\dagger} + \mbox{const.}
\label{e7}
\ee
with
\begin{eqnarray}
t_{\alpha,k}^\dagger & = & u_k \gamma_{\alpha,k}^\dagger+v_k \gamma_{
\alpha,-k}^{\phantom\dagger},
\label{e6}\\
\omega_k  & = & J_1 \sqrt{1+2 e_k}\label{e8}
\\
e_k & = &\frac{J_2}{J_1} \cos{k}
\label{e9}\\
u[v]_k^2
& = & \frac{1}{2}\left(\frac{J_1 (1+ e_k)}{\omega_k}+[-]1\right)
\label{e10}
\end{eqnarray}
The '$[]$'-bracketed sign on the rhs. in (\ref{e10}) refers to the
quantity $v$ on the lhs.. The spin-gap $\Delta=\min\{\omega_k\}$ resides
at $k=\pi$ with $\Delta=\sqrt{J_1^2-2J_1J_2}$.  Note that because of
(\ref{e6}) the ground state $|D\rangle$, which is defined by
$\gamma_{\alpha,k}|D\rangle=0$, contains quantum-fluctuations beyond the
pure singlet product-state. To leading order the dispersion $\omega_k$ is
identical to perturbative expansions\cite{Barnes93,Reigrotzki94}. Beyond
the LHP approach triplet interactions and more elaborate consideration of
the constraint (\ref{e5}) lead to a renormalization of $\omega_k$ and the
formation of multi-magnon bound
states\cite{jurecka00-2,Kotov98,Kotov98-2,Kotov99,jurecka01-2}. However,
for $J_2\ll J_1$ these renormalizations can be neglected.
\end{section}
 
\begin{section}{Two-hole Hamiltonian}
\label{sec4}
In the two hole sector a mapping of the Hamiltonian (\ref{e1}) onto a rung
basis requires for operators which describe singly occupied as well as empty
rungs in addition to the singlet and triplet operators (\ref{e4}) of
the undoped spin system. We label a rung occupied by a single-hole by
introducing an additional pseudo-fermion\cite{jurecka01} (holon). I.e.
instead of the rung being in one of the states given by (\ref{e4}) it can
also be in the states
\begin{eqnarray}
a_{j,n,\sigma}^\dagger|0\rangle=|j\sigma\rangle_n \label{e11a}
\end{eqnarray}
where the l.h.s. denotes the vacuum, i.e. $|0\rangle$, with a single rung
at site $n$ in a one-hole state of spin $\sigma$ with $j=1,2$ referring to
the two positions available for the hole on the bond. Moreover two holes on
the same rung are created by an additional boson
\begin{eqnarray}
b_{n}^\dagger|0\rangle=|00\rangle_n \label{e11b},
\end{eqnarray}
where now the l.h.s. refers to the vacuum $|0\rangle$ with one hole-pair
at site $n$. Both, the single-hole operators $a_{j,i,\sigma}$ and the
hole-pair operators $b_n$ are required to obey the standard fermionic
and bosonic operator algebras.

In the two-hole sector each rung necessarily is in one of the states
(\ref{e4}), (\ref{e11a}) or (\ref{e11b}). As a consequence an extended
hard-core constraint (\ref{e5}) has to be fulfilled
\be
s_n^\dagger s_n^{\phantom\dagger}+t_{\alpha,n}^\dagger t_{
\alpha,n}^{\phantom\dagger}+\sum_{j=1,2} a_{j,n, \sigma}^\dagger
a_{j,n, \sigma}^{\phantom\dagger}+b_n^\dagger b_n^{\phantom\dagger}=1
\label{e13a}
\ee
with a summation over repeated spin-indices implied.

Creation of a physical hole in the (half-filled) ground state $|D\rangle$
of the spin-system  is described by\cite{park01}
\begin{eqnarray}\label{e13}
\hat{c}_{j, n, \sigma} =&& \frac{p_j}{\sqrt{2}}[
a_{\overline{j}, n, \overline{\sigma}}^\dagger
(p_j p_\sigma s_n^{\phantom\dagger} + t_{z,n}^{\phantom\dagger})
\nonumber\\
&& +a_{\overline{j}, n, \sigma}^\dagger
(p_{\overline{\sigma}} t_{x,n}^{\phantom\dagger}
+i t_{y,n})^{\phantom\dagger}]+b^\dagger_n a_{j,n,\sigma}
\end{eqnarray}
where $p_j=+(-)$, $\overline{j}=2(1)$ for $j=1(2)$ and $p_\sigma= +(-)$,
$\overline{\sigma}=\downarrow(\uparrow)$ for $\sigma=\uparrow
(\downarrow)$.  In terms of (\ref{e13}) the creation of a
physical hole can be interpreted as the removal of a spin-rung followed
by the creation of a single-hole state or the removal of a single-hole state
followed by the creation of a hole-pair state. The particular linear
combination of the $a_{j, n,\sigma}^\dagger$, $s_n^{\phantom\dagger}$, and
$t_{\alpha,n}^{\phantom\dagger}$ operators on the r.h.s. ensures that the
total spin is $S=1/2$ and $S_z=\pm 1/2$. Using the constraint (\ref{e13a})
and the bosonic (fermionic) commutation relations for the $b_n$
($a_{j,n,\sigma}$) it is straightforward to show, that the r.h.s. of
(\ref{e13}) indeed satisfies the usual Hubbard-operator algebra. The 
pseudo-particle representation of the spin operators reads
\begin{eqnarray}
S^\alpha_{j,n}&=&\frac{1}{2}(p_j s_n^\dagger t_{\alpha, n}^{}
+p_j t_{\alpha, n}^\dagger  s_n^{} -
i\sum_{\beta, \gamma} \epsilon_{\alpha\beta\gamma} 
t_n^\dagger t_\gamma^{})\nonumber \\&&+\frac{1}{2}\sum_{\sigma,\sigma'} 
a_{j, n, \sigma}^\dagger \tau^\alpha_{\sigma\sigma'} 
a_{j, n, \sigma'}^{}\label{e13bbb}.
\end{eqnarray}
where $\tau^\alpha_{\sigma,\sigma'}$ are Pauli-matrices.

Inserting (\ref{e13}) and (\ref{e13bbb}) into Hamiltonian (\ref{e1})
various processes are found: First, single-hole processes like (i)
intra-rung hopping, (ii) inter-rung hopping, and (iii) exchange scattering
of the pseudo-fermions. Second, two-hole processes result like the (iv) 
decay of a hole-pair on a rung and a singlet (triplet) into two 
separated pseudo-fermions, (v) exchange scattering between two pseudo-fermions
 on neighboring rungs and
(vi) an on-site binding energy of the hole-pair, due to the additional
singlet in the system as compared to a situation with two holes occupying
separate rungs. Finally, three-hole processes occur which involve one
hole-pair and one pseudo-fermion creation operator. These can be discarded
in the two-hole sector. Processes (i) - (iii) constitute the Hamiltonian
$H_1$ of the {\em one}-hole sector which has been discussed in 
ref.\cite{jurecka01}. For the sake of completeness $H_1$ is listed in
appendix \ref{apa}. The remaining processes (v) and (vi)
are contained in $H_b$ while $H_t$ represents process (iv) of the
the final Hamiltonian which reads
\begin{eqnarray}
H&=&H_1+H_{t}+H_{J}\label{e16}\\
H_t&=&\frac{t_2s}{\sqrt{2}}\sum_{n,j,\sigma} p_\sigma b^\dagger_{n+1} a_{j,n,\sigma} a_{\overline{j},n+1,\overline{\sigma}}\nonumber \\ &&\qquad\qquad+ |n\leftrightarrow n+1| +h.c. \nonumber \\
&-&\frac{t_2}{\sqrt{2}}\sum_{n,j,\sigma} b^\dagger_{n+1}[ t^\dagger_{z,n} a_{\overline{j},n,\sigma}\nonumber \\&&\qquad\qquad+(p_\sigma t^\dagger_{x,n}-i t^\dagger_{y,n}) a_{\overline{j},n,\overline{\sigma}}] a_{j,n+1,\overline{\sigma}}\nonumber \\
&&\qquad\qquad+|n\leftrightarrow n+1| +h.c.
\label{e17} \\
H_b&=&-J_1 \sum_n b^\dagger_n b^{\phantom\dagger}_n\nonumber\\
\label{e17a}
&+&J_2 \sum_{n, j} ({\bf S}^{jj}_{n,n}{\bf S}^{jj}_{n+1,n+1}-\frac{1}{4}n^a_{n,j}n^a_{n+1,j}), 
\end{eqnarray}
where $\overline{j}=2(1)$ for $j=1(2)$ and $p_\sigma=+(-)$,
$\overline{\sigma}=\downarrow(\uparrow)$ for $\sigma=\uparrow
(\downarrow)$. ${\bf S}_{m,n}^{ij}=\frac{1}{2}
\sum_{\sigma_1,\sigma_2}a^\dagger_{i,m,\sigma_1}{\bf
\tau}_{\sigma_1,\sigma_2}a^{\phantom{\dagger}}_{j,n,\sigma_2}$ and
$n^a_{n,j}$ is the number operator of fermion $a_{j,n}$. Following the
LHP approximation we have replaced the singlet operators by $C$-numbers,
i.e. $s$.  Within the LHP approach $s=1$ to lowest order. Regarding the
charge dynamics we improve upon this by requiring that the condensate
density of the singlet is determined by satisfying the hard-core
constraint (\ref{e5}) on the average, i.e. by setting
$s^{\phantom\dagger}_n=s^\dagger_n=\langle s_n\rangle=s$ with
\be
s^2=1-\sum_\alpha \langle t_{\alpha, n}^{\dagger}
t_{\alpha, n}^{\phantom\dagger} \rangle =1-\frac{3}{N}\sum_q v_q^2.
\label{e14}
\ee
In principle this equation can be used to determine the LHP
magnon-dispersion selfconsistently which however we will refrain
from here.

After Fourier transforming
(\ref{e16}-\ref{e17a}) and inserting the Bogoliubov representation
(\ref{e6}) we get
\begin{eqnarray}
H&=&H_1+H_S+H_T+H_b \label{e18} \\
H_S&=&\sum_{k,q}S_{kq}b^\dagger_k\times\nonumber\\ \label{e18a}
&\times& (a^{\phantom\dagger}_{0,k-q,\downarrow}a^{\phantom\dagger}_{0,q,\uparrow}+a^{\phantom\dagger}_{\pi,k-q,\uparrow}a^{\phantom\dagger}_{\pi,q,\downarrow})+h.c.\\
H_T&=&\sum_{k,q,q'} T_{kqq'}b^\dagger_{k-q}\times\nonumber\\ \label{e18b}
&\times&\left [ \right . t_{q, z}^\dagger (a^{\phantom\dagger}_{\pi,q',\uparrow}a^{\phantom\dagger}_{0,k-q',\downarrow}+a^{\phantom\dagger}_{\pi,q',\downarrow}a^{\phantom\dagger}_{0,k-q',\uparrow})\nonumber \\&&+t_{q, x}^\dagger (a^{\phantom\dagger}_{0,q',\uparrow}a^{\phantom\dagger}_{\pi,k-q',\uparrow}-a^{\phantom\dagger}_{0,q',\downarrow}a^{\phantom\dagger}_{\pi,k-q',\downarrow})\nonumber \\&&+i t_{q, y}^\dagger (a^{\phantom\dagger}_{0,q',\uparrow}a^{\phantom\dagger}_{\pi,k-q',\uparrow}+a^{\phantom\dagger}_{0,q',\downarrow}a^{\phantom\dagger}_{\pi,k-q',\downarrow}) \left.\right]+h.c.\\
H_b&=&-J_1\sum_k b^\dagger_k b^{\phantom\dagger}_k\label{e18c}
\end{eqnarray} 
with
\begin{eqnarray}
S_{kq}&=&\frac{\sqrt{2} t_2 s}{\sqrt{N}} \left[\cos{(k-q)}+\cos{(q)}\right]\label{e19}\\
T_{kqq'}&=&\frac{\sqrt{2}t_2}{N}\left[\cos{(q-q')}+\cos{(k-q-q')}\right]\label{e20}
\end{eqnarray}
and the (anti)bonding linear combinations of single-hole states
\be
a^\dagger_{0(\pi),n,\sigma}=\frac{1}{\sqrt{2}}\left(a_{2,n,\sigma}^\dagger\pm
a_{1,n,\sigma}^\dagger\right)
\ee
which have momentum $0(\pi)$ i.e. even (odd) parity in rung direction, where
the $+(-)$ belongs to $0(\pi)$. In (\ref{e18c}), and as we focus on the limit
of $J_2\ll J_1$, where the LHP provides for a controlled approximation
of the spin system, we have dropped the exchange scattering $\propto J_2$
from $H_b$ of (\ref{e17a}).

In the two-hole sector the hard-core constraint (\ref{e13a}) is of
particular importance in order to suppress double occupancies of a
single rung by two $a$-type pseudo-fermions. This cannot be achieved by
enforcing (\ref{e13a}) only on the average, rather an auxiliary intra-rung
hard-core potential has to be added to the Hamiltonian (\ref{e18}-\ref{e18c})
\be
U=U_0\sum_{n,\sigma,\sigma',j,j'}
a^\dagger_{j,n,\sigma}a^\dagger_{j',n,\sigma'}a^{\phantom\dagger}_{j',n,\sigma'}a^{\phantom\dagger}_{j,n,\sigma}
\label{e22}
\ee
with $U_0\rightarrow \infty$. A similar hard-core potential, however
to study the two-{\em triplet} sector has been used in refs.
\cite{Kotov98,Kotov98-2,Kotov99}.

\end{section}

\begin{section}{Two-hole spectrum}
\label{sec5}

The primary goal of this work is to evaluate the momentum dependent energy
spectrum of a {\em two}-hole state on the ladder. While the physical {\em
single}-hole excitations are composite objects, c.f. (\ref{e13}), their
deviations from the pseudo fermions are small for $J_2\ll J_1$, i.e., for
a small triplet density induced by quantum fluctuations (see also
ref.\cite{jurecka01}). Therefore, as a first step towards our goal we
approximate the two-hole energy spectrum by that of two pseudo-fermions
based on the limit of strong intra-rung exchange. This leaves (i) a 
two-particle problem to be solved, with however (ii) strongly renormalized
one-particle excitations. On the bare two-particle level (i) can be achieved
{\em exactly} by summing the T-matrix incorporating the interactions $H_S$,
$H_T$ and $U$. For an approximate account of (ii), we will sum the T-matrix
using renormalized one-particle pseudo-fermion Green's functions
\be
G_{k_y,\sigma}(k,t)=-i \Theta(t)\langle D|\{a_{k_y, k, \sigma}(t) 
a_{k_y, k,\sigma}^\dagger\}|D\rangle.\label{e22a}
\ee
The main source of renormalization for a single pseudo-fermion is
known to be multi-triplet emission which can be accounted for by
a selfconsistent resummation of all noncrossing
n-loop graphs to the self energy\cite{jurecka01}. This is equivalent to the 
selfconsistent-Born-approximation (SCBA) which has been employed for the
one- and two-dimensional
$t$-$J$-model\cite{Schmitt-Rink88,Kane89,Martinez91,Liu92}. 
On the ladder the SCBA leaves a {\em finite} quasi-particle residue to the
pseudo-fermion propagator, which is a consequence of the spin gap.
Yet, the spectra acquire a substantial incoherent part as well as
renormalization of the quasi-particle dispersion (for details
see ref.\cite{jurecka01}).

To evaluate the T-matrix we note that the two-leg ladder is symmetric with
respect to reflections at a plane perpendicular to the rungs\cite{gopalan94}.
This leaves the corresponding parity a good quantum number. The
the (anti)bonding single-hole states $a^\dagger_{(\pi)0,n,\sigma}|D\rangle$
are of even(odd) parity, the hole-pair state $b^\dagger_n|D\rangle$ is
of even parity, and
the triplet $t^\dagger_{\alpha, n}|D\rangle $ is of odd parity. Apart from
parity the two-hole states can be classified additionally according to
their total spin, with four subspaces, i.e., $S=0$ even, $S=0$
odd, $S=1$ even, and $S=1$ odd arising. Two-pseudo-fermion scattering
induced by Hamiltonian (\ref{e18}-\ref{e18c}) occurs only in the the $S=0$
even channel (via $H_S$) and in the $S=1$ odd channel (via $H_T$).
Therefore we will focus on these channels in the following.

\begin{subsection}{$S=0$ even channel}

\begin{figure}
\centerline{\psfig{file=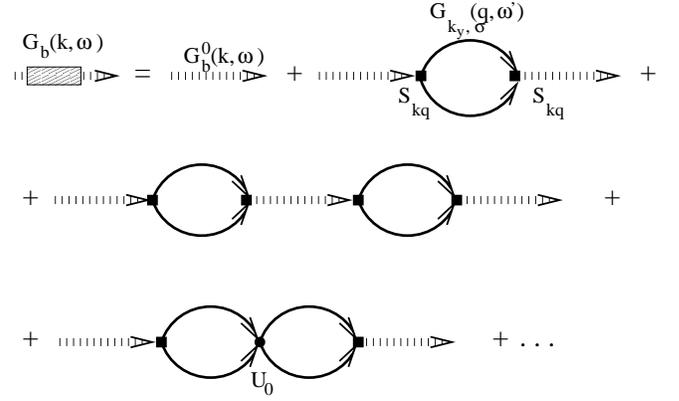,width=\linewidth,angle=-90}}   
\vspace{0.5cm}
\caption{T-matrix hole-pair Green's function $G_b(k,\omega)$. Squares: scattering vertex $S_{kq}$ (\ref{e18a}). Circles: hard-core repulsion $U_0$ (\ref{e22}). Full lines: SCBA pseudo-fermion Green's functions. Dotted lines: bare hole-pair Green's functions. }
\label{fig2}
\end{figure}

To calculate the even parity $S=0$ energy spectrum we start from the
hole-pair Green's function
\be
G_b(k,t)=-i\Theta(t)\langle D|[b_k^{\phantom\dagger}(t)
b^\dagger_k]|D\rangle ,
\label{e23}
\ee
where the bare hole-pair propagator is set by $H_b$ of (\ref{e18c})
\be
G_b^0(\omega)=\frac{1}{\omega+J_1}
\label{e24}
\ee
and for the remainder of this work retarded propagators with
$\omega\equiv\omega+i\eta$ and $\eta\rightarrow0^+$ are implied. 
Following the preceding discussion the interacting propagator
$G_b(k,\omega)$ is calculated by the T-matrix summation of
fig. \ref{fig2}
\be
G_b(k,\omega) =\pmatrix{
0 &  0 & G_b^0(\omega)}(1_3-M(k,\omega))^{-1}\pmatrix{0\cr0\cr 1}
\label{e26}
\ee
where $M(k,\omega)$ is a $3\times3$ matrix with
\begin{eqnarray}
M_{11(22)}(k,\omega)&=&\frac{U_0}{N} \sum_q I_{00(\pi\pi)}(k,q,\omega)\nonumber \\
M_{12(21)}(k,\omega)&=&0\nonumber \\
M_{13(23)}(k,\omega)&=&\sqrt{\frac{U_0}{N}}G_b^0(\omega) \sum_q S_{kq}I_{00(\pi\pi)}(k,q,\omega)\nonumber \\
M_{31(32)}(k,\omega)&=&\sqrt{\frac{U_0}{N}}\sum_q S_{kq} I_{00(\pi\pi)}(k,q,\omega) \nonumber \\
M_{33}(k,\omega)&=& G_b^0(\omega)\sum_{q,a=0,\pi} S_{kq}^2 I_{aa}(k,q,\omega)
\end{eqnarray}
with $S_{kq}$ from (\ref{e24}) and $U_0$ from (\ref{e22}).
$I_{\mu\nu}(k,q,\omega)$ refers to the two-pseudo-fermion bubble
\begin{eqnarray}
I_{\mu\nu}(k,q,\omega)=-\frac{1}{\pi}\int d\omega' 
(&&\mbox{Im}G_{\mu,\sigma}(q, \omega'))
\nonumber\\
&&G_{\nu,\sigma}(k-q,\omega-\omega')
\label{e28}
\end{eqnarray}
with $\mu=0,\pi$, $\nu=0,\pi$ and the SCBA renormalized pseudo-fermion Green's functions
$G_{k_y,\sigma}(k,\omega)$ of (\ref{e22a}). On the l.h.s. of (\ref{e28}) the
spin index $\sigma$ has been omitted since $G_{k_y,\sigma}(k,\omega)$ and
thus $I_{\mu\nu}(k,\omega)$ is spin-independent. This implies that,
while collective excitations may appear only in the $S=0$ even and the $S=1$ odd
subspace, the two-pseudo-fermion {\em continua} are degenerate with respect
to total spin. For even parity states only diagonal elements of $I_{\mu\nu}$
enter the calculation, odd parity states imply off-diagonal elements.

\end{subsection}

\begin{subsection}{$S=1$ odd channel}
The odd parity part of the spectrum is extracted from the triplet-hole-pair 
Green's function
\be
G_{bt}(k,\omega)=-i\Theta(t) \sum_q g_{kq}^2 \langle
D|b_{k-q}^{\phantom\dagger} \gamma_{z,q}^{\phantom\dagger}
\gamma_{z,q}^\dagger b_{k-q}^\dagger |D\rangle
\label{e36}
\ee
where $\gamma_{z,q}$ is the Bogoliubov-transform (\ref{e6}) of the
triplet and the choice of the $z$ triplet is arbitrary due to spin
rotational invariance. The form factor
$g_{kq}=(2/N)^{1/2}u_q\cos(\frac{k}{2}-q)$ has been introduced
for computational convenience: the matrix elements of $H_T$
which enter the T-matrix resummation are of the form $u_qu_{q'}T_{kqq'}=
2\sqrt{2}t_2 g_{kq}g_{kq'}$ where, due to the limit $J_1\gg J_2$, we have
neglected contributions $\propto v_q$. Thus
$g_{kq}$ reduces the number of distinct
two-particle reducible element to the T-matrix summation. This form factor
changes the weight but not the position of the T-matrix eigenvalues.

The hole-pair-triplet Green's function is coupled to two pseudo-fermion
scattering states via $H_T$. With respect to the interactions
$H_T$ and $U$ the T-matrix evaluation, fig. \ref{fig3}, proceeds
analogous to that in $S=0$ even subspace. Note however, that to 
the zeroth order in the pseudo-fermion scattering states the
hole-pair-triplet Green's function $G_{bt}^0(k,\omega)$ 
\begin{eqnarray}
G_{bt}^0&&(k,\omega)=-\frac{1}{\pi}\sum_q g_{kq}^2\int d\omega'
(\mbox{Im}D(q,\omega') \times
\nonumber \\
&& G_b(k-q,\omega-\omega') =\sum_q
g_{kq}^2G_b(k-q,\omega-\omega_q)
\label{e37}
\end{eqnarray}
contains interaction effects already, since $G_b(k,\omega)$ is the {\em 
interacting} hole-pair Green's function (\ref{e26}) of Fig. \ref{fig2}.
$D(q,\omega)=1/(\omega-\omega_q)$ is the triplet Green's function
obtained from (\ref{e7}). The full T-matrix reads

\begin{figure}
\centerline{\psfig{file=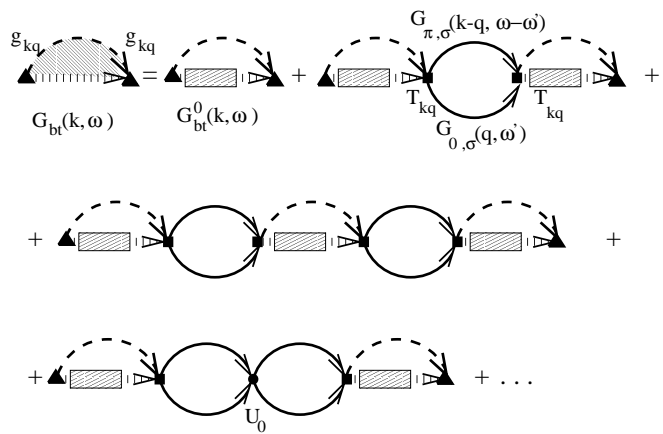,width=\linewidth,angle=-90}}   
\vspace{0.5cm}
\caption{T-matrix hole-pair-triplet Green's function $G_{bt}({k,\omega})$. Squares: Scattering vertex $T_{kqq'}$ (\ref{e18b}). Circles: hard-core repulsion $U_0$ (\ref{e22}). Dashed line: Triplet Green's function. Full line: SCBA pseudo-fermion Green's function. Dotted line with box: T-matrix hole-pair Green's function of fig. \ref{fig2}.}
\label{fig3}
\end{figure}

\be
G_{bt}(k,\omega) =\pmatrix{
0 & G_{bt}^0(k,\omega)}(1_2-M'(k,\omega))^{-1}\pmatrix{0\cr 1}
\label{e38}
\ee
with
the $2\times 2$ matrix 
\begin{eqnarray}
M'_{11}&=&\frac{U_0}{N}\sum_q I_{0\pi}(k, q, \omega)\nonumber \\
M'_{12}&=&\frac{2\sqrt{U_0}t_2}{\sqrt{N}}G_{bt}^0(k, \omega)\sum_q g_{kq} I_{0\pi}(k, q, \omega)\nonumber\\
M'_{21}&=&\frac{2\sqrt{U_0}t_2}{\sqrt{N}}\sum_q g_{kq} I_{0\pi}(k, q, \omega)\nonumber\\
M'_{22}&=&4t_2^2 G_{bt}^0(k, \omega)\sum_q g_{kq}^2 I_{0\pi}(k ,q, \omega). \label{e39}
\end{eqnarray}

\end{subsection}
\end{section}
\begin{section}{Results and Discussion}
\label{sec7}

\begin{figure}
\centerline{\psfig{file=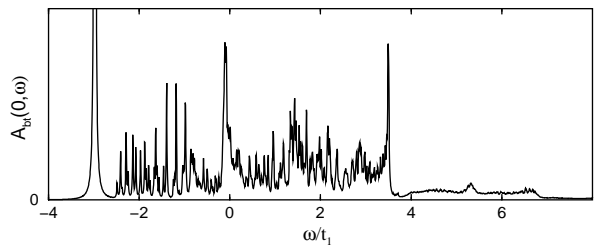,width=0.9\linewidth,angle=-90}}   
\caption{Hole-pair-triplet spectral function 
$A_{bt}(k,\omega)$ at 
$k=0$. $J_1/t_1=3$, $J_2/t_1=0.3$, $t_1=t_2=1$, $N=32$.}
\label{fig5a}
\end{figure}

The energy spectrum in the even (odd) parity sector of the two-hole spectrum
results from an evaluation of the hole-pair-(triplet) Green's functions
$G_{b(bt)}(k,\omega)$ and a subsequent determination of the regions of
nonvanishing spectral
density $A_{b(bt)}(k,\omega)=-\mbox{Im}G_{b(bt)}(k,\omega)/\pi$,
an example of which is depicted in Fig. \ref{fig5a} for
$k=0$ at $J_1/t_1=3$, $J_2/t_1=0.3$, $t_1=t_2=1$, and $N=32$.
In evaluating $G_{b(bt)}$, both the $\omega$
integrations and $k$ summations are performed numerically for small, but
finite $\eta$ and $U_0/t_1=10^6$. We have checked that beyond the latter
value the spectrum is insensitive to any further increase of $U_0$. Moreover,
and except for later comparison to results of exact diagonalization we
use a system size of $N=32$ rungs for the remainder of this work. We have
checked that for larger system sizes the spectrum remains almost unchanged
suggesting that our results are sufficiently close to the thermodynamic
limit.

\begin{figure}
\centerline{\psfig{file=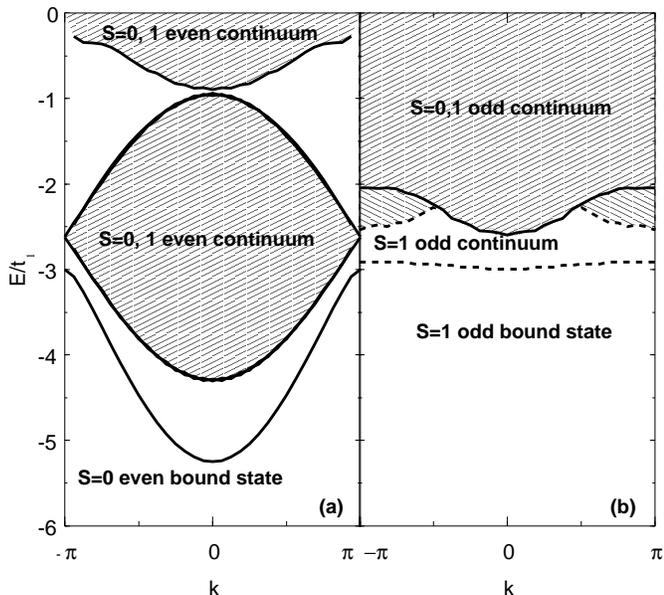,width=\linewidth}}   
\caption{Low lying two-hole excitations of a spin ladder with $N=32$ rungs and $J_1/t_1=3$, $J_2/t_1=0.3$ and $t_1=t_2$. (a) even parity (b) odd parity.}
\label{fig5}
\end{figure}

Figure~\ref{fig5}a (b) shows the dispersion of the even (odd) parity low
energy two-hole
excitations of a spin ladder with $J_2/t_1=3$, $J_1/t_1=0.3$ and 
$t_2=t_1$.  The state of lowest energy is a two-hole bound state with $S=0$
and even parity, which is split off from the continuum. The binding effect
is due to two holes residing on the same rung  and thereby
maximizing the number of singlets\cite{dagotto92}. The overall
ground state of the ladder is this $S=0$ even bound-state at $k=0$. The
charge excitations of lowest energy are intra-band excitations in the $S=0$ even bound state and therefore gapless, while the lowest spin excitations are inter-band excitations between the ground state and the $S=1$ even continuum and 
therefore are gapful. 
Thus a Luther-Emery\cite{Luther74} rather than a
Luttinger-liquid\cite{Haldane80} applies to the present case. This ground
state is consistent with that found by Lanczos
diagonalization\cite{tsunetsugu94,troyer96}, density matrix
renormalization group (DMRG)\cite{white97}, and high order
series expansion\cite{oitmaa99}.

In the odd parity spectrum of fig. {\ref{fig5}}b an $S=1$ bound state is
observed which is consistent with results from exact 
diagonalization\cite{troyer96,dagotto98,riera99,poilblanc00}.
In the T-matrix this state arises as a collective excitation due to local
mixing of the hole-pair-triplet state with two pseudo-fermions.
For very strong coupling $J_1\gg J_2, t_1, t_2$ an additional antibound
state\cite{laeuchli99} arises, which however merges with the
continuum already at intermediate couplings, because the continuum
is strongly spread due to the incoherent part of the single-hole spectra.

\begin{figure}
\centerline{\psfig{file=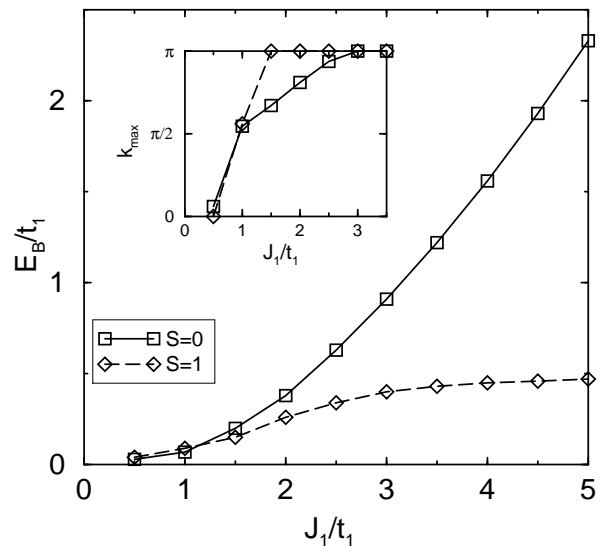,width=0.9\linewidth,angle=-90}}   
\caption{Bound state binding energy $E_B$ for $S=0$ (squares) 
and $S=1$ (diamonds). $J_2/J_1=0.1$, $t_1=t_2$, $N=32$. Inset:
 maximum $k$ at which the bound state exists. 
}
\label{fig8}
\end{figure}

To emphasize the difference between the bound states in the $S=0$ and
$S=1$ sector, we show the binding energy $E_B\equiv{\rm min}
(E_{\rm cont})-{\rm min}(E_{\rm bs})$ as a function of $J_1$ for a constant
ratio $J_2/J_1=0.1$ in fig. \ref{fig8} where
${\rm min}(E_{{\rm cont}({\rm bs})})$
denotes the minimum energy of the continuum (bound state).
 For large $J_1$ a linear behavior of the 
$S=0$ binding
energy is found in contrast to a constant binding energy for the $S=1$
state. This difference reflects the different binding
mechanisms: the energy gain of the $S=0$ bound state 
is due to the additional singlet which accompanies this
state\cite{dagotto92}. In this case the energy gain is $\propto J_1$.
In the $S=1$ sector, and regarding $J_1$ only, the local hole-pair-triplet
state is energetically equivalent to a two pseudo-fermion scattering
state. Therefore binding in this channel is mainly due to kinetic
delocalization of the hole-pair-triplet into two pseudo-fermions
on the neighboring sites\cite{laeuchli99}. We note, that
only for a finite range of parameters $J_1/t_1\gtrsim 1$ the bound states
exist over
the entire Brillouin zone. This is shown in the inset of fig. \ref{fig8}
which depicts the maximum values of $k$ for which the bound states 
can be separated from the continuum.

Returning to fig.~\ref{fig5}(a) and (b) it is evident, that in the
even parity sector the high energy states form two continua, degenerate in
$S=0$ and $S=1$. The lower one of these, i.e. between $E\approx-4 t_1$
and $E\approx-t_1$, is due to two pseudo-fermion
scattering states formed out of the bonding orbitals, while the higher
ones are due to either the incoherent parts of the bonding spectra or
to two pseudo-fermion scattering states both in antibonding orbitals.
Note that in fig. \ref{fig5} we have cut off the spectrum at high
energies with additional continua existing above the cut off.

The lower bound of the $S=1$ odd continuum in fig.\ref{fig5}(b) shows two
local minima at $k=0$ and $k=\pi$. This structure is due to the two types
of excitations which contribute to the continuum\cite{lin98}: (i)
scattering states
of a triplet and the hole-pair on a rung, i.e. the hole-pair-triplet bubble
in fig. \ref{fig3} of the
diagrammatic theory, and (ii) the continuum of two pseudo-fermions, i.e.
the pseudo-fermion bubble in fig. \ref{fig3}. The latter is degenerate
with the $S=0$ odd
excitations and its minimum of energy is at $k=\pi$, while the former
one has no counterpart with $S=0$ and has its minimum at $k=0$. This
excitation has a dispersion of its lower edge which is reminiscent of
that of the single triplet on the undoped ladder. For the
parameters chosen in Fig. (\ref{fig5}) both minima of the continua
are nearly degenerate
whereas in general it depends on the particular choice of
$t_1,t_2,J_1,J_2$, which minimum is at the lowest energy.

\begin{figure}
\centerline{\psfig{file=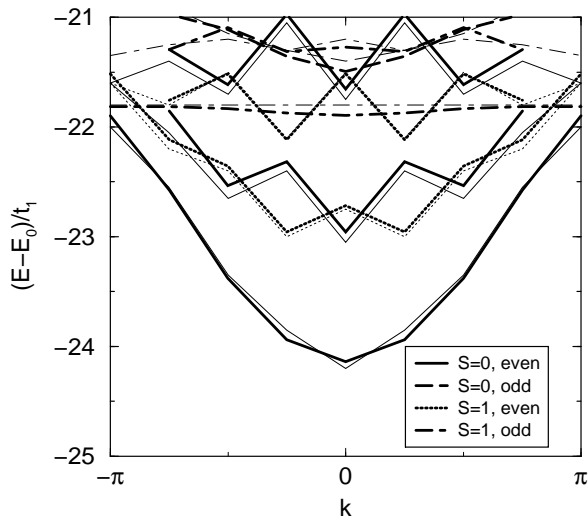,width=0.9\linewidth,angle=-90}}   
\caption{Low lying two-hole excitations of a spin ladder with $N=8$ and $J_1/t_1=3$, $J_2/t_1=0.3$ and $t_1=t_2$. Thin lines: reproduced after Troyer {\it et al.}$^8$. The perturbative spectrum has been shifted by a constant ground state energy of $E_0=-19.1 t_1$.}
\label{fig6}
\end{figure}

Next a comparison of our diagrammatic results with exact diagonalization
data is performed. To this end, in fig. \ref{fig6}, a two-hole spectrum
reproduced from the work of Troyer et al. \cite{troyer96} is depicted
together with our results from a T-matrix evaluation. In order to perform
the comparison identical parameters, and in particular identical systems
sizes have been chosen. Thick lines in this figure represent 
the diagrammatic, thin lines the numerical result. Obviously the agreement 
is very good. We emphasize that this kind of agreement is promoted by 
the limit of strong rung coupling $J_1/J_2\gg 1$ which pertains to fig.
\ref{fig6}. We do not expect similar agreement for $J_2/J_1\rightarrow 1$.

\begin{figure}[bt]
\centerline{\psfig{file=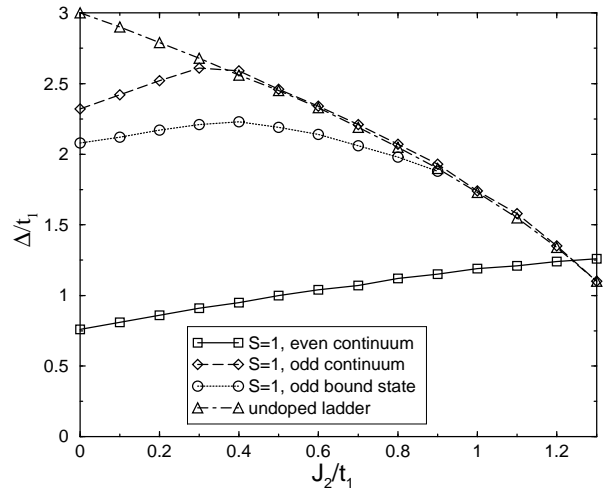,width=0.9\linewidth,angle=-90}}   
\caption{Excitation energies of the different $S=1$ states as function 
of $J_2$. $J_1/t_1=3.0$, $t_1=t_2$, $N=32$.}
\label{fig9}
\end{figure}

To conclude this section we discuss the spin-gap of the doped ladder.
From the spectrum of fig. \ref{fig5} it is obvious that various $S=1$
states are found in the doped ladder\cite{lin98}: (i) the even parity
continuum, (ii) the odd parity bound state, and (iii) the odd parity
continuum
consisting (a) of triplet-hole-pair scattering states with a minimum at
$k=\pi$ and (b) of two pseudo-fermion scattering states with minimum at
$k=0$. All of these excitations are gapped, with the actual spin gap set
by the lowest energy $S=1$ excitation. Fig. \ref{fig9} shows the
excitation energies for all of the different $S=1$ states for $J_1/t_1=3$
and $t_1=t_2$ as a function of $J_2$. The spin gap of the undoped ladder
is also displayed for comparison.
For nearly the entire range of $J_2$ in fig. \ref{fig9} the even parity
continuum (i) defines the spin gap.
The gap-size is set by the energy difference to the $S=0$ bound state
at $k=0$.
The continuum (i) has no counterpart in the undoped ladder. Thus we find
that the
spin gap evolves discontinuously upon doping in this parameter
regime. The excitations at higher energies are the bound state (ii) and 
the continuum (iii)
for $J_2/t_1<0.9$, while the bound states merges with the continuum at
this particular value of $J_2$. For $J_2>0.3$ the lower edge of the odd
parity continuum
is set by the triplet-hole-pair scattering state, i.e. (iii a). As noted
above this edge of the continuum has a dispersion
similar to that of the triplet excitation on the undoped ladder and
therefore leads to a spin gap also comparable to that of the undoped ladder.
For
$J_2/t_1<0.3$ the lowest odd parity continuum state is the two
pseudo-fermion state, i.e. (iii b), which explains the deviation from
the spin gap of the undoped ladder. Finally at
$J_2/t_1\approx 1.2$ a level crossing occurs and the hole-pair-triplet
continuum, i.e. (iii b), becomes the lowest $S=1$ excitation. Only in this
parameter regime the spin gap evolves continuously upon doping.

Unfortunately the LHP approximation considerably underestimates the
spin-gap of the triplet excitation for $J_2\rightarrow J_1$ compared 
to numerically exact results \cite{greven96}.  Therefore
the results for the spin-gap in Fig. \ref{fig9} are qualitative only
as $J_2$ increases. Nevertheless the level crossing cited above has
been found also in numerical diagonalization\cite{riera99}, however at a
parameter range not accessible to our diagrammatic approach. Introducing a
diagonal next-near-neighbor hopping $t'$ between site $1(2)$ 
of rung $n$ and site $2(1)$ of rung $n+1$ it has been shown in numerical
studies\cite{poilblanc00} that the spin gap can be tuned continuously
from being set by the triplet of the undoped ladder via the
hole-pair-triplet bound-state to the even parity continuum. Finally,
very recently, it was found that a strongly negative $t'$ changes the
nature of the ground state from the
$S=0$ bound state to that of the $S=0$ even continuum\cite{tsutsui01},
thus from a
Luther-Emery liquid\cite{Luther74} to a scenario with gapless charge
and spin excitations.
\end{section}

\begin{section}{Conclusions}
In conclusion we have analyzed the excitations in spin ladders doped with
two holes. Applying a mapping of the generalized $t$-$J$ model onto a
coupled boson-fermion model the impact of both, scattering by the spin
background and hole-hole interactions on the two hole spectrum has been
analyzed in detail. We find the ground state to be a two-hole bound state,
whereas
excitations consist of various continuum states with $S=0,1$ and even or
odd parity. Additionally, depending on the particular choice of parameter,
$S=1$ odd parity bound and antibound states are found. The binding energy
of the bound states and the spin gap of the system have been analyzed. Our
results compare well with numerical studies of finite systems. Yet,
our method allows for a study of system sizes close to the thermodynamic
limit.
\end{section}

\begin{acknowledgements}
This research was supported in part by the Deutsche Forschungsgemeinschaft
under Grant No. BR 1084/1-1 and BR 1084/1-2.
\end{acknowledgements}

\appendix
\section{}\label{apa} 
In this appendix, and for the sake of completeness we list the
Hamiltonian $H_1$ of (\ref{e16}), i.e. of the single-hole sector
\begin{eqnarray}
H_1&=&-t_1 \sum_{n, \sigma} a_{1,n,\sigma}^\dagger
a_{2,n,\sigma}^{\phantom\dagger} + h.c. \nonumber \\
&+&\frac{t_2}{2} s^2 \sum_{j, n, \sigma} a_{j, n, \sigma}^\dagger a_{j,
n-1, \sigma}^{\phantom\dagger}+h.c. \nonumber \\
&+&t_2 s \sum_{j, n}{\bf t}_n^\dagger (-1)^{j-1} \left({\bf S}_{n-1,
n}^{jj}+{\bf S}_{n+1,n}^{jj}\right)+h.c. \nonumber\\
&+& \frac{J_2}{2} s \sum_{j, n}{\bf t}_n^\dagger (-1)^{j-1} 
\left({\bf S}_{n-1,n-1}^{jj}+{\bf S}_{n+1,
n+1}^{jj}\right)+h.c. \nonumber\\
&+&\frac{t_2}{2} \sum_{j, n, \sigma} {\bf t}_{n-1}^\dagger {\bf
t}_{n}^{\phantom\dagger} a_{j, n, \sigma}^\dagger a_{j, n-1,
\sigma}^{\phantom\dagger}+h.c. \nonumber \\
&-&t_2 \sum_{j, n} i \left ({\bf t}_n^\dagger \times {\bf
t}_{n+1}^{\phantom\dagger} \right ) {\bf S}_{n+1, n}^{jj} 
+h.c. \nonumber \\
&-&\frac{J_2}{2} \sum_{j, n} i \left ({\bf t}_n^\dagger \times {\bf
t}_{n}^{\phantom\dagger} \right ) \left ({\bf S}_{n+1, n+1}^{jj}
+{\bf S}_{n-1,
n-1}^{jj}\right). \label{eA1}
\end{eqnarray}
where
${\bf t}_n^\dagger=(t_{x,n}^\dagger,t_{y,n}^\dagger, t_{z,n}^\dagger)$.
This Hamiltonian includes (i) intra-rung hopping, 
(ii) inter-dimer hopping
and (iii) exchange scattering.
 Interdimer hopping can be either spin-diagonal
or accompanied by spin-flip scattering. This includes (ii,a) singlet-singlet
and (ii,b) singlet-triplet and (ii,c) triplet-triplet transitions of the spin 
background upon hole doping. For further details see Ref. \cite{jurecka01}.

\end{document}